\documentclass[aps,prl,twocolumn,floatfix,nofootinbib,10pt]{revtex4-2}
\usepackage{graphicx,amsmath,amssymb,bm,xcolor,orcidlink,float}
\hypersetup{colorlinks=true,linkcolor=blue,citecolor=blue,urlcolor=blue}
\begin{document}

\title{Reabsorbed Urca neutrinos suppress bulk-viscous damping of neutron star merger oscillations}

\author{Prashant Thakur~\orcidlink{0000-0003-4189-6176}$^{*}$}
\author{Yeunhwan Lim~\orcidlink{0000-0001-9252-823X}$^{\dagger}$}
\affiliation{Department of Physics, Yonsei University, Seoul 03722, South Korea}

\date{\today}

\begin{abstract}
Reabsorption of Urca neutrinos suppresses the bulk-viscous damping predicted for neutron star merger oscillations. Below the direct-Urca threshold the capture neutrinos have mean energies of 19--46~MeV and are absorbed within metres. An energy-resolved kinetic treatment gives 2.7~s for a 1-kHz, 10-km oscillation at nuclear density and 4~MeV, against 19~ms for freely escaping neutrinos. It holds across the sampled states, three equations of state and the rate variations, with nearly frozen composition. This absorption is present in full-kinematics rates and absent from elastic opacity tables.
\end{abstract}

\maketitle
\renewcommand{\thefootnote}{\fnsymbol{footnote}}\footnotetext[1]{prashant@yonsei.ac.kr}\footnotetext[2]{Contact author: ylim@yonsei.ac.kr}\renewcommand{\thefootnote}{\arabic{footnote}}

\textit{Introduction.}---The remnant of a binary neutron star merger oscillates at kilohertz frequencies and radiates gravitational waves for tens of milliseconds, and their spectrum encodes the equation of state (EOS)~\cite{Bauswein:2011tp,Takami:2014zpa}. Urca bulk viscosity has been proposed to damp these oscillations during the signal: the composition lags behind the density oscillation, and the lag dissipates energy most efficiently where the Urca relaxation rate matches the mode frequency, at a few MeV~\cite{Alford:2017rxf,Alford:2018lhf,Alford:2019qtm,Most:2021zvc,Alford:2021ogv,Alford:2023gxq}. Simulations and forecasts with this viscosity suggest possible waveform imprints~\cite{Most:2022yhe,Chabanov:2023blf,Zhu:2026gpx}, whereas simulations with neutrino transport find no significant imprint~\cite{Zappa:2022rpd} or only a brief window for bulk-viscous effects~\cite{Espino:2023dei}. The transparent bulk viscosity assumes that the Urca neutrinos escape, on the criterion that a neutrino with energy equal to the temperature has a scattering length of kilometres below about 5~MeV~\cite{Alford:2018lhf,Harris:2020rus,Alford:2021ogv}. At higher temperatures the neutrinos are taken to be trapped, with a bulk viscosity smaller by orders of magnitude~\cite{Alford:2019kdw}, and simulations switch between the regimes at a cutoff temperature~\cite{Most:2022yhe}. The neutrinos emitted by electron capture, however, carry energies well above the temperature. Below the direct-Urca threshold their scale is set by the momentum deficit of the reaction, about 15~MeV at twice nuclear density, and detailed balance ties their absorption to their emission. Charged-current rates with the full kinematics are available~\cite{Reddy:1997yr,Roberts:2016mwj,Sokolowski:2026pub,Jaeger:2026lnh,Alford:2024xfb,Tambe:2026slj}, whereas the elastic approximation of the opacity tables keeps the nucleon momentum fixed~\cite{Bruenn:1985en,Burrows:2004vq}. Recent work extends the transparent viscosity to muons, $\Delta$ isobars and superfluid matter~\cite{Hernandez:2026tak,Kumar:2026uvs,Zhang:2026csf}, and transport studies compute neutrino distributions in merger remnants~\cite{Alford:2026kwd,Ng:2026zrd}. Neutrino transport in dense matter has long been studied~\cite{Goodwin:1982hy}, and in newly born neutron stars neutrino absorption was found to change the weak-interaction damping of radial pulsations completely, giving damping times of the order of seconds~\cite{Sawyer:1980wp}. Below the direct-Urca threshold, however, no calculation has connected the energy-dependent emission, reabsorption and spatial transport of the Urca neutrinos to the damping of merger oscillations. Here we take this step with an energy-resolved kinetic treatment. The capture neutrinos are reabsorbed within metres, and a 1-kHz, 10-km oscillation at nuclear density and 4~MeV damps in 2.7~s, compared with 19~ms for the transparent baseline with the same rates and 25--100~ms on the published map~\cite{Alford:2019qtm}. The millisecond window disappears at every sampled density and temperature and for all three EOSs. For these global oscillations, Urca damping at a few MeV, where it was expected to act, takes far longer than the tens-of-milliseconds gravitational-wave signal.

\textit{Methodology.}---We consider the electron-Urca sector of $npe\mu$ matter described by the DD2 EOS~\cite{Typel:2009sy}, with the cold $\beta$-equilibrium composition, the muon fraction held fixed, and the resulting finite-temperature affinity $\Delta\mu_0=\mu_n-\mu_p-\mu_e$. Direct Urca is evaluated as a momentum-resolved phase-space integral with a constant weak vertex and Landau effective masses, a simplification of the full relativistic rates of Ref.~\cite{Alford:2021ogv}. At the energies that dominate the capture spectrum it agrees with the degenerate open-threshold result of Ref.~\cite{Reddy:1997yr} and with exact-dispersion phase-space integrations to about 10\% (see Supplemental Material~\cite{SupplementalMaterial}). Modified Urca is taken in the Fermi-surface approximation~\cite{Friman:1979ecl}. Neutral-current scattering on neutrons and protons is treated as isotropic and elastic, with its total rate taken from the full nonrelativistic response~\cite{Reddy:1997yr}. The absorption length of a $\nu_e$ of energy $E$ follows from the emission spectrum by detailed balance. With rates per unit time and volume,
\begin{equation}
\frac{1}{\lambda_\nu(E)}=\frac{2\pi^2(\hbar c)^3}{cE^2}\,e^{(E+\Delta\mu_0)/T}\,\frac{d\Gamma^{(0)}_{\rm cap}}{dE},
\label{eq:db}
\end{equation}
where $\Gamma^{(0)}_{\rm cap}$ is the capture rate with the final-neutrino blocking factor set to unity. The same relation with $\Delta\mu_0\to-\Delta\mu_0$ and the decay spectrum gives the $\bar\nu_e$ absorption length.

The response to a density oscillation is obtained from a linearised kinetic treatment of the $\nu_e$ and $\bar\nu_e$ populations coupled to the matter. We linearise around the occupancies $f_i^0(E)$, $i=\nu,\bar\nu$, in equilibrium with the matter at chemical potentials $\mp\Delta\mu_0$. A compression $\epsilon=\delta n/n$ perturbs the affinity by $\delta\mu=C\epsilon-Bq$, where $q=n\,\delta Y_e$ is the composition change produced by the reactions and
\begin{equation}
C=n\Big(\frac{\partial\Delta\mu}{\partial n}\Big)_{T,Y_e,Y_\mu},\qquad B=-\Big(\frac{\partial\Delta\mu}{\partial n_e}\Big)_{n,T,Y_\mu}>0
\label{eq:CB}
\end{equation}
follow from the finite-temperature EOS. Each neutrino mode relaxes toward the shifted equilibrium $\delta f_{{\rm eq},i}=\mp f_i^0(1-f_i^0)\,\delta\mu/T$ at the stimulated rate $\gamma_i(E)=c/[\lambda_i(E)(1-f_i^0)]$, with the emission fixed by detailed balance, and is scattered isotropically at the rate $\sigma(E)=c/\lambda_{\rm NC}(E)$. For a plane wave $e^{i(kx-\omega t)}$ the linearised kinetic equation reads
\begin{equation}
(\gamma_i+\sigma-i\omega+ick\mu)\,\delta f_i=\gamma_i\,\delta f_{{\rm eq},i}+\sigma\langle\delta f_i\rangle,
\label{eq:kin}
\end{equation}
with $\mu$ the direction cosine. Its angular average is $\langle\delta f_i\rangle=\gamma_iA_i\,\delta f_{{\rm eq},i}/(1-\sigma A_i)$ with $A_i(E)=\arctan[ck/(\gamma_i+\sigma-i\omega)]/(ck)$. The net lepton exchange per unit affinity, summed over both species, defines the admittance
\begin{equation}
\Lambda(\omega,k)=\sum_i\int dE\,w_i\gamma_i\left[1-\frac{\gamma_iA_i}{1-\sigma A_i}\right],
\label{eq:lam}
\end{equation}
with the spectral weight $w_i=E^2f_i^0(1-f_i^0)/[2\pi^2T(\hbar c)^3]$, so that $dq/dt=\Lambda\,\delta\mu$. The cycle-averaged dissipated power per volume is ${\rm Re}(\delta\mu^*\Lambda\,\delta\mu)/2$, which defines an effective, wavenumber-dependent bulk viscosity and the damping time of the oscillation energy,
\begin{equation}
\zeta_{\rm eff}(\omega,k)=\frac{C^2\,{\rm Re}\,\Lambda}{|{-i\omega}+B\Lambda|^2},\qquad \tau=\frac{K_T}{\zeta_{\rm eff}\,\omega^2},
\label{eq:zeta}
\end{equation}
with $K_T=n(\partial p/\partial n)_{T,Y_e,Y_\mu}$. The amplitude damping time is $2\tau$.

Two limits of Eq.~(\ref{eq:zeta}) are used below. Without spatial transport, $k\to0$, the scattering drops out and $\Lambda_i=\int dE\,w_i\gamma_i(-i\omega)/(\gamma_i-i\omega)$, the closed-channel response of the trapped-neutrino regime~\cite{Sawyer:1980wp,Alford:2019kdw}. In the short-wavelength limit $A_i\to0$ and $\Lambda\to\Lambda_\infty=\sum_i\int dE\,w_i\gamma_i$, a real response coefficient, and $\zeta_{\rm eff}$ takes the transparent form $C^2\Lambda_\infty/(\omega^2+B^2\Lambda_\infty^2)$ with relaxation rate $B\Lambda_\infty$. The transparent baseline follows Refs.~\cite{Alford:2018lhf,Alford:2019qtm,Alford:2023gxq} with the same rates. All neutrinos escape, and the electron fraction relaxes to the value $Y_e^*$ at which decay and capture balance. With the net rate $F=\Gamma_{\rm dec}-\Gamma_{\rm cap}$, the baseline viscosity is $\zeta_{\rm tr}=nC\gamma_B/(\omega^2+\gamma_I^2)$, where $\gamma_I=-n^{-1}\partial F/\partial Y_e$ and $\gamma_B=\partial F/\partial n$, and all quantities are evaluated at $Y_e^*$. Unlike the short-wavelength limit of Eq.~(\ref{eq:zeta}), the baseline has no neutrinos and a different composition. The treatment is linear in the oscillation amplitude, isothermal, local in a stationary background, and restricted to the electron sector with elastic scattering. Muonic reactions, inelastic scattering and thermal driving of the rates are left for future work.

\begin{figure}[t]
\includegraphics[width=\columnwidth]{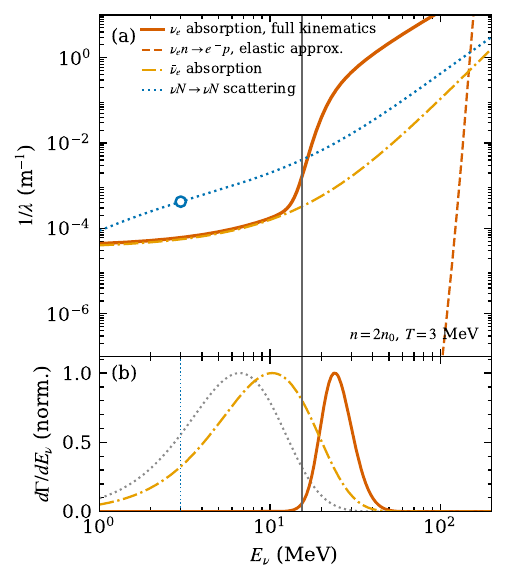}
\caption{(a) Inverse mean free path of neutrinos in DD2 matter at $n=2n_0$, $T=3$~MeV: $\nu_e$ absorption by direct and modified Urca (solid), direct Urca in the elastic approximation (dashed), $\bar\nu_e$ absorption by inverse modified Urca (dash-dotted), and neutral-current scattering (dotted). The circle marks the $E_\nu=T$ scattering length of the transparency criterion~\cite{Alford:2018lhf}. The vertical line marks the kinematic scale $E_{\min}$ of the capture spectrum. (b) Normalised emission spectra: electron capture (solid), modified-Urca $\bar\nu_e$ (dash-dotted), and $E^2f(E)$ (grey dotted).}
\label{fig:mfp}
\end{figure}

\textit{Results.}---Figure~\ref{fig:mfp}(a) compares the mean free path on which the transparency argument rests with the absorption length of the neutrinos that are actually emitted, at $2n_0$ and $T=3$~MeV. For a neutrino with energy $E=T$ the neutral-current scattering length is 2.4~km and the absorption length 16~km, the basis of the usual estimate that merger matter below $T\approx5$~MeV is transparent~\cite{Alford:2018lhf}. For a 25-MeV neutrino the absorption length is 5~m. The relevant energy is set by the kinematics of electron capture below the direct-Urca threshold, where particles confined to their Fermi surfaces cannot satisfy the kinematic constraints of $e^-+p\to n+\nu_e$: the momentum deficit $\Delta p=p_{Fn}-p_{Fp}-p_{Fe}$ is 49~MeV at $2n_0$ (momenta in energy units). As discussed in Ref.~\cite{Alford:2018lhf}, the deficit is covered most cheaply by a proton excited above its Fermi surface, which gains momentum $\epsilon_p/v_{Fp}$ at an energy cost $\epsilon_p$ ($v_{Fp}=0.46$ at $2n_0$, in units of $c$), together with the neutrino momentum itself, directed opposite to that of the neutron: $\Delta p\simeq\epsilon_p/v_{Fp}+E$ with $E\simeq\epsilon_p$. This introduces a kinematic scale well above $T$,
\begin{equation}
E_{\min}\simeq\frac{v_{Fp}\,\Delta p}{1+v_{Fp}},
\label{eq:emin}
\end{equation}
15~MeV at $2n_0$. Figure~\ref{fig:mfp}(b) shows the resulting direct-capture number spectrum: it peaks around 24~MeV with a mean energy of 27~MeV, and 0.5\% of it lies below $E_{\min}$. By Eq.~(\ref{eq:db}) the absorption length of these neutrinos is tied to their emission: at the dominant capture energies the inverse reaction is not suppressed, because it returns the proton to the same state above the Fermi sea, which is the origin of the 5-m absorption length in Fig.~\ref{fig:mfp}(a). The same holds across the sampled DD2 states from $0.5$ to $3n_0$ and 2 to 8~MeV, where the mean capture energy is 19--46~MeV, the emission-weighted harmonic mean absorption length stays below 7~m, and the probability that a capture neutrino travels 1~km is at most 3.1\%. Opacity tables built in the elastic approximation~\cite{Bruenn:1985en,Burrows:2004vq} keep the nucleon momentum fixed and do not include this proton-excitation route: the final electron is Pauli blocked unless $E\gtrsim\mu_e-Q=161$~MeV at $2n_0$ (131~MeV with the mean-field shift of Ref.~\cite{Oertel:2020pcg}), so at the dominant capture energies, far below this crossover, elastic absorption is negligible and the dashed curve of Fig.~\ref{fig:mfp}(a) rises into the panel only above 100~MeV, whereas full-kinematics rates include this route~\cite{Roberts:2016mwj,Jaeger:2026lnh}. The antineutrinos from modified-Urca decay are absorbed mainly through the inverse modified-Urca reaction, within a few kilometres at this state and over much shorter distances at higher temperature. The energy-resolved treatment of Eqs.~(\ref{eq:kin})--(\ref{eq:zeta}) includes whatever escape remains, which matters for the antineutrinos at the lowest temperatures. For the direct-capture neutrinos, therefore, the free-escape assumption behind the transparent bulk viscosity does not hold: they are reabsorbed within metres of their emission, far inside the wavelength of any global mode, and the capture channel exchanges lepton number with a local neutrino population rather than carrying it out of the fluid element.

\begin{figure}[t]
\includegraphics[width=\columnwidth]{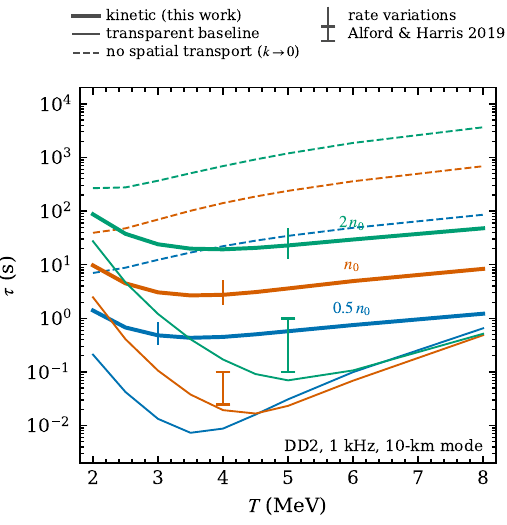}
\caption{Damping time versus temperature for a 1~kHz, 10-km density oscillation in DD2 matter at $0.5n_0$, $n_0$, and $2n_0$: kinetic result (thick), this work's transparent baseline (thin), and the limit without spatial transport, $k\to0$ (dashed). Thin vertical bars show the range of damping times from varying Urca and scattering rates. Capped bars show the intervals read from the contour map of Ref.~\cite{Alford:2019qtm} at the same physical density.}
\label{fig:tau}
\end{figure}

Figure~\ref{fig:tau} shows the consequence for the damping of a 1-kHz, 10-km density oscillation, Eq.~(\ref{eq:zeta}), at $0.5n_0$, $n_0$ and $2n_0$ between 2 and 8~MeV. The transparent baseline, computed with the same rates as the kinetic result, has the familiar resonance where the Urca relaxation rate matches the mode frequency: it falls to 7.3~ms at $0.5n_0$ and 3.5~MeV, 17~ms at $n_0$ and 4.5~MeV, and 70~ms at $2n_0$ and 5~MeV, and varies by a factor of 89 to 390 across the temperature range. The kinetic result has no such window. It stays between 0.43 and 1.4~s at $0.5n_0$, 2.7 and 9.7~s at $n_0$, and 19 and 87~s at $2n_0$, with shallow minima of 0.43, 2.7 and 19~s near 3.5--4~MeV, and at the five reference states of Table~\ref{tab:tau}, chosen near the transparent minima, it is longer by factors of 36 to 490. The two sets of curves approach each other only where the transparent time is itself long: 0.2--27~s at 2~MeV, and 0.65~s at $0.5n_0$ and 8~MeV.

The dashed curves, the $k\to0$ limit of Eq.~(\ref{eq:lam}), show where the difference comes from. At the Table~\ref{tab:tau} states even the slowest of the sampled neutrino modes between $T$ and $10T$ relaxes to the local matter within 5--27~$\mu$s, well inside the 1-ms period, and a population that follows the matter dissipates little: without spatial transport the damping time lies a further factor of 26 to 52 above the kinetic result and lengthens with temperature as the rates increase. Most of the dissipation that remains comes from the transport of lepton number across the mode by the neutrinos themselves, and at the Table~\ref{tab:tau} states the antineutrinos account for about 60--80\% of the dissipative response. This transport is hindered by neutral-current scattering: doubling the scattering rate lengthens the $n_0$ damping time from 2.74 to 4.33~s. Faster Urca reactions do not shorten it: at the Table~\ref{tab:tau} states from $n_0$ upward, tripling the modified-Urca rate or doubling the direct-Urca rate lengthens the kinetic damping time, at $n_0$ to 3.36 and 2.86~s.

\begin{table*}[t]
\caption{Damping time of a 1~kHz, 10-km density oscillation in DD2 matter at reference states near the transparent damping minima. The transparent result of this work, $\tau_{\rm tr}$, is compared with the reading from the contour map of Ref.~\cite{Alford:2019qtm} and with the kinetic result $\tau_{\rm kin}$ including neutrino reabsorption. The range reflects rescaled Urca and scattering rates. The last column gives $\tau_{\rm kin}$ for the IUFSU and BigApple EOSs. Dashes mark entries outside the map or near the direct-Urca threshold.}
\label{tab:tau}
\begin{ruledtabular}
\begin{tabular}{lcccccc}
$n/n_0$ & $T$ (MeV) & $\tau_{\rm tr}$ (ms) & Ref.~\cite{Alford:2019qtm} (ms) & $\tau_{\rm kin}$ (s) & range (s) & IUFSU / BigApple (s)\\
\hline
0.5 & 3 & 13.3 & -- & 0.48 & 0.32--0.86 & 0.56 / 0.27\\
1   & 4 & 19.4 & 25--100 & 2.74 & 1.8--5.2 & 8.4 / 64.4\\
1.5 & 4 & 70.2 & 50--100 & 8.65 & 5.6--15.9 & 4.9 / 4.4\\
2   & 5 & 69.7 & 100--1000 & 22.9 & 13.0--47.8 & 7.7 / 7.4\\
3   & 5 & 198 & $10^3$--$10^4$ & 96.7 & 54--202 & -- / --\\
\end{tabular}
\end{ruledtabular}
\end{table*}

The marks in Fig.~\ref{fig:tau} are readings of the published contour map of Ref.~\cite{Alford:2019qtm}, which used a different treatment of the rates. At the same physical density it gives 25--100~ms at $n_0$ and 0.1--1~s at $2n_0$, and it does not extend down to $0.5n_0$. Against the published interval at $n_0$ and 4~MeV the kinetic damping time is 27 to 110 times longer, and against the baseline with the same rates 141 times longer. Table~\ref{tab:tau} collects the five states. The range column comes from varying the modified-Urca rate by a factor of three and the direct-Urca and scattering rates by a factor of two in either direction. The damping times stay within a factor of about two of the central values, and even their lower ends lie 24 to 270 times above the transparent ones. The IUFSU~\cite{Fattoyev:2010mx} and BigApple~\cite{Fattoyev:2020cws} EOSs, treated in the same way and with their published nucleon mass of 939~MeV, give 0.27--64~s at the same states, 85 to 500 times longer than their transparent baselines of 2.2--130~ms. The entries at $3n_0$ are omitted because both EOSs lie within 16~MeV of the direct-Urca threshold there, where the rate treatment is not validated. A second implementation, with exact quasiparticle dispersions and a direct angular integration, reproduces the damping times at the DD2 states of Table~\ref{tab:tau} to within 2\% (see Supplemental Material~\cite{SupplementalMaterial}).

\begin{figure}[t]
\includegraphics[width=\columnwidth]{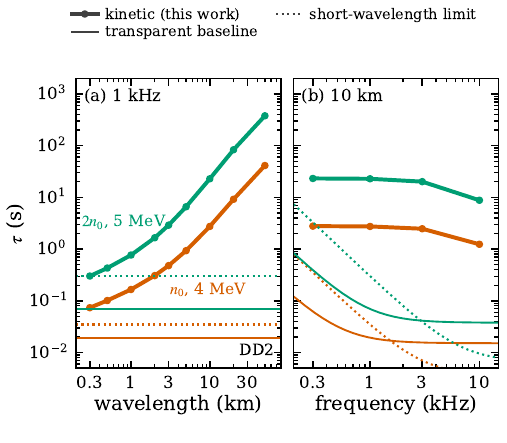}
\caption{Damping time of a density oscillation in DD2 matter at $n_0$, 4~MeV and $2n_0$, 5~MeV: (a) versus mode wavelength at 1~kHz and (b) versus frequency for a 10-km mode. Thick curves with dots give the kinetic result at the computed points. Thin solid lines show the transparent baseline and dotted lines the short-wavelength limit of the kinetic model, at 1~kHz in (a) and evaluated at each frequency in (b).}
\label{fig:scope}
\end{figure}

Figure~\ref{fig:scope} shows how the damping depends on the wavelength and frequency of the mode. In panel (a) the kinetic damping time at 1~kHz rises steadily with the wavelength: at $n_0$ and 4~MeV from 74~ms at 0.3~km to 2.7~s at 10~km and 41~s at 50~km, and at $2n_0$ and 5~MeV from 0.30~s to 23~s and 377~s. The effective viscosity depends on the wavenumber as well as on the local state. Only structure below a kilometre approaches the short-wavelength limit of the model (dotted), in which the neutrinos cross the compressed region before they interact and $\zeta_{\rm eff}$ takes the transparent form: 35~ms and 0.30~s at the two states. This limit is 1.8 and 4.3 times the transparent baseline (thin), which assumes that all neutrinos escape and uses the stationary composition. Panel (b) shows the weak dependence on the frequency. For the 10-km mode at the two states shown, the kinetic damping time changes by less than 14\% between 0.3 and 3~kHz and is shorter by a factor of 2 to 3 at 10~kHz, whereas between 0.3 and 10~kHz the transparent baseline falls by a factor of 4 to 10 and the short-wavelength limit by a factor of 90 to 340. Between 1 and 3~kHz the change over the whole sampled grid stays below 14\% at $n\ge n_0$ and below 12\% at $0.5n_0$. The seconds of Fig.~\ref{fig:tau} and Table~\ref{tab:tau} are therefore not tied to the 1-kHz, 10-km reference: they lengthen for longer wavelengths and change little across the kHz band.

For merger simulations the practical consequence is the composition response. For a small-amplitude 1-kHz, 10-km mode the amplitude of the electron-fraction oscillation per MeV of affinity, $|\Lambda|/(\omega n)$, is $4.0\times10^{-6}$ at $n_0$ and 4~MeV, against the fully relaxed static reference $1/(Bn)=8.9\times10^{-4}$. Over the sampled states from $0.5$ to $3n_0$ and 3 to 5~MeV the reduction is a factor of 72 to 725. On the oscillation timescale the composition is nearly frozen, and the small out-of-phase part of the response provides the dissipation. The response includes the spatial transport of lepton number across the mode and describes the oscillation about a fixed background composition. Transport calculations with full-kinematics charged-current rates find the core absorption opacity of $\nu_e$ larger by many orders of magnitude than in the elastic approximation, favouring trapping~\cite{Jaeger:2026lnh}, and recover trapped weak equilibrium in the core~\cite{Ng:2026zrd}, while a Monte Carlo study with elastic opacities finds the neutrino distributions approaching free streaming in the cooler regions of a merger snapshot~\cite{Alford:2026kwd}. These findings are consistent with the absorption lengths of Fig.~\ref{fig:mfp}(a) for the DD2 states considered here. The present calculation adds the step from the emitted spectrum to the damping of the oscillation.

\textit{Conclusions.}---Below the direct-Urca threshold the capture neutrinos have mean energies of 19 to 46~MeV and are reabsorbed within metres, so the transparent Urca bulk viscosity does not apply to a merger core at these temperatures: a 10-km, 1-kHz mode at $n_0$ and 4~MeV damps in 2.7~s instead of 19~ms, and in at least 0.43~s over the sampled DD2 grid. The result follows from detailed balance, which ties the absorption of the capture neutrinos to their emission, and it holds under the tested rate variations, for three EOSs and across the kHz band. For the kHz oscillations of a remnant the composition is nearly frozen on the oscillation timescale, so the Urca channel damps them on seconds rather than tens of milliseconds. Full-kinematics transport rates contain the proton excitation responsible for this reabsorption, a channel absent from elastic opacity tables.

\begin{acknowledgments}
P.~T. and Y.~L. are supported by a National Research Foundation of Korea (NRF) grant funded by the Korean government (MSIT) (No.~RS-2024-00457037). P.~T. was also supported in part by the Yonsei University Research Fund (Yonsei University Frontier Fellowship for Postdoctoral Researchers) in 2025. Y.~L. acknowledges support from the Global--Learning \& Academic Research Institution for Master's and PhD Students, and Postdocs (LAMP) Program through a National Research Foundation of Korea (NRF) grant funded by the Ministry of Education (No.~RS-2024-00442483). The code and the result files that reproduce every figure and table of this Letter are available at \url{https://github.com/prashantstar123/urca-neutrino-reabsorption}.
\end{acknowledgments}

\bibliography{refs}

\onecolumngrid
\clearpage
\setcounter{secnumdepth}{3}
\setcounter{section}{0}
\setcounter{equation}{0}
\setcounter{figure}{0}
\setcounter{table}{0}
\renewcommand{\thesection}{S\arabic{section}}
\renewcommand{\theequation}{S\arabic{equation}}
\renewcommand{\thefigure}{S\arabic{figure}}
\renewcommand{\thetable}{S\arabic{table}}
\begin{center}
{\large\bfseries Supplemental Material for: Reabsorbed Urca neutrinos suppress bulk-viscous damping of neutron star merger oscillations\par}
\vspace{10pt}
Prashant Thakur and Yeunhwan Lim\par
\vspace{4pt}
{\small\itshape Department of Physics, Yonsei University, Seoul 03722, South Korea\par}
\end{center}
\vspace{14pt}

\section{Equation of state, background state and thermodynamic coefficients}
\label{sec:S1}

The reference calculations use the DD2 relativistic mean-field EOS~\cite{Typel:2009sy} for neutrons, protons, electrons and muons, with saturation density $n_0=0.149$~fm$^{-3}$. At each baryon density $n$ the composition is fixed at its cold $\beta$-equilibrium value, which gives the electron and muon fractions $Y_e$ and $Y_\mu$ listed in Table~\ref{tab:S1}, and the same composition is used at every temperature. The calculation describes the electron-Urca sector. Muonic reactions are not included, and the muon fraction is held fixed during the oscillation. At temperature $T$ this composition has the affinity $\Delta\mu_0=\mu_n-\mu_p-\mu_e$, obtained from the finite-temperature state at fixed composition, and the neutrino populations are in equilibrium with the matter at the chemical potentials $-\Delta\mu_0$ for $\nu_e$ and $+\Delta\mu_0$ for $\bar\nu_e$. The reaction rates use the nucleon Dirac effective masses $m^*_n$ and $m^*_p$ of the cold background, through the Landau masses $m_L=(p_F^2+m^{*2})^{1/2}$.

The response to a density oscillation involves three thermodynamic coefficients, defined in Eqs.~(2) and (5) of the main text,
\begin{equation}
C=n\left(\frac{\partial\Delta\mu}{\partial n}\right)_{T,Y_e,Y_\mu},\qquad
B=-\left(\frac{\partial\Delta\mu}{\partial n_e}\right)_{n,T,Y_\mu},\qquad
K_T=n\left(\frac{\partial p}{\partial n}\right)_{T,Y_e,Y_\mu}.
\label{eq:S1}
\end{equation}
They are evaluated at the temperature of each state by central finite differences of the finite-temperature state, with a relative step of $2\times10^{-3}$. Halving the step changes $C$ and $B$ by less than $10^{-6}$ in relative terms. As an independent check, the relation $C=-(\partial p/\partial n_e)_{n,T,Y_\mu}$, which follows from the free energy at fixed temperature, holds to a relative accuracy of $1.2\times10^{-6}$ at all five states. The transparent baseline of the main text evaluates $C$ and $K_T$ in the same way, but at its own stationary electron fraction $Y_e^*$.

Below the direct-Urca threshold the Fermi momenta satisfy $p_{Fn}>p_{Fp}+p_{Fe}$, and the momentum deficit $\Delta p=p_{Fn}-p_{Fp}-p_{Fe}$ is 34--89~MeV at the five states. The momentum deficit is most economically supplied by a proton excited above its Fermi surface together with the emitted neutrino momentum~\cite{Alford:2018lhf}, and this sets the energy scale of the emitted neutrinos. With the proton excitation energy linearised, this gives Eq.~(6) of the main text, $E_{\min}\simeq v_{Fp}\Delta p/(1+v_{Fp})$. With the exact relativistic proton dispersion,
\begin{equation}
E_{\min}=\left[(p_{Fp}+\Delta p-E_{\min})^2+m_p^{*2}\right]^{1/2}-\left(p_{Fp}^2+m_p^{*2}\right)^{1/2},
\label{eq:S2}
\end{equation}
the scale is 13.1--16.7~MeV at the five states. Because the dispersion is convex, the linear estimate lies below the exact value, by 1.5\% at $3n_0$ and by 28\% at $0.5n_0$, where the proton has to move far from its Fermi surface. Equations~(6) and (\ref{eq:S2}) define cold, zero-affinity reference scales, with the electron and the neutron at their Fermi surfaces. The finite-temperature rate calculations use the listed affinity, and their spectrum has a low-energy tail. At $2n_0$ and 3~MeV, Fig.~1(b) of the main text places about 0.5\% of the capture emission below the linear scale of Eq.~(6).

\begin{table}[H]
\caption{Background state and thermodynamic coefficients of DD2 matter at the five reference states of Table~I of the main text. Momenta, masses, energies, $\Delta\mu_0$ and $C$ in MeV, $n$ in fm$^{-3}$, $B$ in MeV\,fm$^3$ and $K_T$ in MeV\,fm$^{-3}$. $E_{\min}$ is given from Eq.~(6) of the main text and from the exact proton dispersion, Eq.~(\ref{eq:S2}), both as cold, zero-affinity reference scales.}
\label{tab:S1}
\begin{ruledtabular}
\begin{tabular}{lccccc}
$n/n_0$ & 0.5 & 1 & 1.5 & 2 & 3\\
$T$ (MeV) & 3 & 4 & 4 & 5 & 5\\
\hline
$n$ & 0.0745 & 0.149 & 0.224 & 0.298 & 0.447\\
$Y_e$ & 0.0329 & 0.0507 & 0.0582 & 0.0633 & 0.0694\\
$Y_\mu$ & 0 & 0.0053 & 0.0181 & 0.0278 & 0.0404\\
$p_{Fn}$ & 254.1 & 317.5 & 360.9 & 395.0 & 449.1\\
$p_{Fp}$ & 82.3 & 123.9 & 157.1 & 183.5 & 223.5\\
$p_{Fe}$ & 82.3 & 119.8 & 143.6 & 162.5 & 191.8\\
$m^*_n$ & 689.5 & 538.1 & 431.9 & 355.5 & 261.5\\
$m^*_p$ & 688.2 & 536.8 & 430.6 & 354.2 & 260.2\\
$v_{Fp}$ & 0.119 & 0.225 & 0.343 & 0.460 & 0.652\\
$\Delta p$ & 89.4 & 73.8 & 60.1 & 49.0 & 33.7\\
$E_{\min}$, Eq.~(6) & 9.5 & 13.5 & 15.4 & 15.4 & 13.3\\
$E_{\min}$, Eq.~(\ref{eq:S2}) & 13.1 & 15.9 & 16.7 & 16.2 & 13.5\\
$\Delta\mu_0$ & 2.10 & 1.32 & 0.72 & 0.81 & 0.59\\
$C$ & 26.4 & 29.4 & 34.5 & 37.9 & 34.9\\
$B$ & 14672 & 7555 & 5487 & 4459 & 3405\\
$K_T$ & 1.05 & 7.55 & 33.1 & 89.2 & 287\\
\end{tabular}
\end{ruledtabular}
\end{table}

\section{Direct-Urca rates}
\label{sec:S2}

Direct-Urca capture $e^-+p\to n+\nu_e$, neutron decay $n\to p+e^-+\bar\nu_e$ and their inverse reactions are evaluated with a constant weak vertex, $\tilde G^2=G_F^2\cos^2\theta_C(1+3g_A^2)$ with $g_A=1.26$, and with the full momentum dependence of the phase space. Each nucleon $j$ is described by the parabolic dispersion $\varepsilon_j(p)=p^2/(2m_{L,j})-\mu_j^T$, with the Landau mass $m_{L,j}=(p_{Fj}^2+m_j^{*2})^{1/2}$, so that the Fermi momenta and Fermi velocities are exact and only the curvature of the dispersion is approximated. The electrons are ultrarelativistic. The energy balance is $\varepsilon_n-\varepsilon_p=E_e\pm E_\nu-\mu_e-\Delta\mu$. The plus sign applies to neutron decay and its inverse, the minus sign to electron capture and its inverse. The nucleon occupations are Fermi--Dirac distributions at the temperature $T$. For each nucleon, $\mu_j^T$ is the chemical potential for which the parabolic dispersion gives the density $n_j$, $n_j=(2m_{L,j}T)^{3/2}F_{1/2}(\mu_j^T/T)/(2\pi^2)$ in units $\hbar=c=1$, where $F_{1/2}(\eta)=\int_0^\infty dx\,x^{1/2}/(1+e^{x-\eta})$ is the complete Fermi--Dirac integral. At zero temperature it reduces to $p_{Fj}^2/(2m_{L,j})$. The electron chemical potential $\mu_e$ and the affinity $\Delta\mu$ are those of the finite-temperature EOS. The angular and nucleon-energy integrals are done analytically, as in Ref.~\cite{Alford:2019kdw} generalised to unequal neutron and proton masses, and the integrals over the electron energy and the momentum transfer are done numerically. All matter occupation factors are kept. The emission rates are computed with the final-neutrino blocking factor set to unity, as in Eq.~(1) of the main text, and the absorption rate per neutrino mode $W(E)$ follows from the inverse reaction.

Emission and absorption are computed from separate integrands, so detailed balance is a check rather than an input. For both species $W(E)$ and the emission spectrum satisfy Eq.~(1) of the main text, with the shift $+\Delta\mu_0$ for $\nu_e$ and $-\Delta\mu_0$ for $\bar\nu_e$. At the five reference states and on the sampled energy grid $0.5T<E<15T$, this identity holds to a relative accuracy of $2\times10^{-8}$ for $\nu_e$, and to $2\times10^{-7}$ for $\bar\nu_e$ wherever direct Urca exceeds $10^{-4}$ of the inverse modified-Urca absorption.

Above the direct-Urca threshold, where the degenerate result of Reddy, Prakash and Lattimer~\cite{Reddy:1997yr} applies, the rate reproduces it to 0.8\% at $E=3$~MeV and $T=3$~MeV, and to 2.6\% and 7.2\% at 10 and 30~MeV. At these energies the degenerate approximation itself loses accuracy: a Monte Carlo integration of the exact phase space, normalised once to the result of Ref.~\cite{Reddy:1997yr} at $E=T=3$~MeV, agrees with our rate to better than 1\% at 10 and 30~MeV. Below the threshold we compare with a separately written implementation. It uses the exact relativistic dispersions with the thermal Dirac masses, the effective chemical potentials of the finite-temperature EOS and massive electrons, and it evaluates the phase-space integrals by direct numerical quadrature. At the five reference states it gives $\nu_e$ absorption rates 0.3--11\% above ours at 20--40~MeV. Weighted with our capture spectrum $S_{\rm cap}$, the ratio of averaged rates,
\begin{equation}
R_{\rm cap}=\frac{\int dE\,S_{\rm cap}(E)\,W_\nu^{\rm exact}(E)}{\int dE\,S_{\rm cap}(E)\,W_\nu^{\rm ours}(E)},
\label{eq:S3}
\end{equation}
is 1.008--1.021 (Table~\ref{tab:S2}). Our $\nu_e$ absorption lengths are therefore slightly conservative. These comparisons are the basis of the statement in the main text that the rate agrees with the open-threshold result and with exact-dispersion integrations to about 10\%.

The second implementation also uses the exact dispersions for the neutral-current rate and integrates the kinetic equation over angles numerically. Used in the kinetic response, it gives damping times within 1.8\% of ours at the five reference states (Table~\ref{tab:S2}), and within 2.6\% at $0.5n_0$ and 4~MeV, at $n_0$ and 8~MeV and at $2n_0$ and 3~MeV. Most of this difference comes from the $\nu_e$ absorption: substituting only the exact $\nu_e$ direct-Urca rate into our calculation lengthens the damping times by 0.5--1.6\%. Antineutrino absorption is dominated by inverse modified Urca (Sec.~\ref{sec:S3}). The direct-Urca fraction of antineutrino absorption, averaged with the weight $w_{\bar\nu}/(\gamma_{\bar\nu}+\sigma)$, is 11\% at $0.5n_0$, 1.6\% at $n_0$ and at most 0.021\% from $1.5n_0$ upward, and substituting only the exact antineutrino direct-Urca rate changes the damping times by less than 0.4\%.

Detailed balance also explains the metre-scale absorption lengths. For the dominant capture route the initial proton lies above its Fermi surface, and the inverse reaction returns it to that state. The factor $e^{(E+\Delta\mu_0)/T}$ in Eq.~(1) then cancels the Boltzmann suppression of the emission, so the absorption at the dominant capture energies is not Boltzmann suppressed. An upper bound on the absorption length follows from restricting the exact-dispersion phase-space integral to configurations with an occupied initial neutron and empty final proton and electron states, where every occupation factor, measured from the finite-temperature chemical potentials, exceeds 1/2. Because the integrand is positive, the restricted integral is a lower bound on the absorption rate. At $2n_0$, 3~MeV and $E=25$~MeV it gives $\lambda\le9.0$~m, or $\lambda\le45$~m if the product of the three occupation factors is replaced by its lower bound 1/8, compared with 5.0~m from our primary calculation (Fig.~1 of the main text).

\begin{table}[H]
\caption{Direct-Urca rates at the five reference states. Columns 3--7: ratio of the $\nu_e$ absorption rate from exact nucleon dispersions to ours, at fixed neutrino energy and as the ratio $R_{\rm cap}$ of capture-spectrum-averaged rates, Eq.~(\ref{eq:S3}). Columns 8--9: $\nu_e$ direct-Urca absorption length at 25~MeV. Columns 10--12: damping times from the two implementations and their relative difference.}
\label{tab:S2}
\begin{ruledtabular}
\begin{tabular}{cccccccccccc}
$n/n_0$ & $T$ & \multicolumn{5}{c}{$W_\nu^{\rm exact}/W_\nu^{\rm ours}$} & \multicolumn{2}{c}{$\lambda_{25}$ (m)} & \multicolumn{3}{c}{$\tau$ (s)}\\
 & (MeV) & 20 & 25 & 30 & 40~MeV & $R_{\rm cap}$ & ours & exact & ours & second & diff.\\
\hline
0.5 & 3 & 1.020 & 1.008 & 1.003 & 1.004 & 1.008 & 2.23 & 2.21 & 0.481 & 0.482 & $+0.16\%$\\
1 & 4 & 1.075 & 1.036 & 1.022 & 1.008 & 1.014 & 3.81 & 3.68 & 2.74 & 2.78 & $+1.41\%$\\
1.5 & 4 & 1.109 & 1.044 & 1.027 & 1.017 & 1.020 & 5.20 & 4.98 & 8.65 & 8.73 & $+0.92\%$\\
2 & 5 & 1.090 & 1.039 & 1.026 & 1.020 & 1.021 & 4.08 & 3.93 & 22.9 & 23.3 & $+1.75\%$\\
3 & 5 & 1.029 & 1.023 & 1.020 & 1.020 & 1.021 & 2.44 & 2.39 & 96.7 & 98.1 & $+1.51\%$\\
\end{tabular}
\end{ruledtabular}
\end{table}

\section{Modified-Urca rates and neutral-current scattering}
\label{sec:S3}

Modified Urca, the reaction $n+N\to p+N+e^-+\bar\nu_e$ with a spectator nucleon $N=n,p$, and its inverse reactions are treated in the Fermi-surface approximation~\cite{Friman:1979ecl}. The rate coefficient $\lambda_{\rm mU}=\partial(\Gamma_{\rm dec}-\Gamma_{\rm cap})/\partial\Delta\mu$, taken at $\Delta\mu=0$, is that of Refs.~\cite{Alford:2019qtm,Alford:2021ogv}. It follows from the one-pion-exchange matrix element, has the numerical prefactors $367/(1152\pi^3)$ for the neutron branch and $367/(8064\pi^3)$ for the proton branch, and is evaluated with the Landau masses of Sec.~\ref{sec:S1}. The energy dependence comes from the phase space of the five degenerate fermions. The emission spectra of the decay antineutrinos and of the capture neutrinos are
\begin{equation}
\frac{d\Gamma_{\rm dec}}{dE}=A\,E^2J_5(E-\Delta\mu),\qquad
\frac{d\Gamma_{\rm cap}}{dE}=A\,E^2J_5(E+\Delta\mu),\qquad
J_5(E)=\frac{E^4+10\pi^2T^2E^2+9\pi^4T^4}{24\,(1+e^{E/T})},
\label{eq:S4}
\end{equation}
and the constant $A$ is fixed by $\lambda_{\rm mU}=4AT^6I_5$, where $I_5=\int_0^\infty x\,J_5(x)\,dx=367\pi^6/6048$ is evaluated at $T=1$. As for direct Urca, the absorption rates follow from the emission spectra by detailed balance. Numerically, $I_5$ is reproduced to $10^{-14}$, and the derivative of the net rate obtained from the spectra reproduces $\lambda_{\rm mU}$ to $5\times10^{-8}$, the accuracy of the finite difference. At $0.5n_0$ the proton Fermi energy, 4.9~MeV, is comparable with the temperature, and the Fermi-surface approximation is less accurate there. The range column of Table~I of the main text includes a variation of $\lambda_{\rm mU}$ by a factor of three in either direction.

Below the direct-Urca threshold, inverse neutron decay, $\bar\nu_e+p+e^-\to n$, is restricted by the momentum mismatch of the Fermi surfaces, whereas in inverse modified Urca the spectator nucleon can supply the missing momentum. Antineutrino absorption is therefore dominated by inverse modified Urca at the reference states from $n_0$ upward (Sec.~\ref{sec:S2}). At $2n_0$ and 3~MeV, the state of Fig.~1 of the main text, the antineutrino absorption length is 18~km at $E=3$~MeV, 6.3~km at 10~MeV and 0.61~km at 30~MeV. Its harmonic mean over the modified-Urca emission spectrum is 2.3~km. At the same density this mean falls to 310~m at 5~MeV and to 48~m at 8~MeV.

Neutral-current scattering on neutrons and protons is computed from the dynamic response of a free nonrelativistic nucleon gas at temperature $T$~\cite{Reddy:1997yr}, with the Landau masses and the chemical potentials $\mu_j^T$ of Sec.~\ref{sec:S2}. In units $\hbar=c=1$, the total scattering rate of a neutrino with momentum $\mathbf p$ and energy $E=|\mathbf p|$ is
\begin{equation}
\frac{1}{\lambda_{\rm NC}(E)}=G_F^2\sum_{j=n,p}\left(h_{V,j}^2+3h_{A,j}^2\right)\int\frac{d^3p'}{(2\pi)^3}\,S_j\big(E-E',|\mathbf p-\mathbf p'|\big),
\label{eq:S5}
\end{equation}
where $E'=|\mathbf p'|$ is the final energy, $h_V=-1/2$ and $h_A=-g_A/2$ for neutrons, and $h_V=1/2-2\sin^2\theta_W$ and $h_A=g_A/2$ for protons, with $\sin^2\theta_W=0.231$. The response of species $j$ to an energy transfer $q_0$ and a momentum transfer $q$ is
\begin{equation}
S_j(q_0,q)=\frac{m_{L,j}^2T}{\pi q}\,\frac{\ln(1+e^{-a})-\ln(1+e^{-a-z})}{1-e^{-z}},\qquad
z=\frac{q_0}{T},\qquad
a=\frac{m_{L,j}}{2q^2T}\left(q_0-\frac{q^2}{2m_{L,j}}\right)^2-\frac{\mu_j^T}{T}.
\label{eq:S6}
\end{equation}
The angular dependence of the weak vertex is averaged, as for the constant vertex of the charged-current rates. Keeping it changes the rate at $2n_0$ and 3~MeV by at most 2.2\% between $E=3$ and 50~MeV. Because the neutrinos here are not degenerate, the rate is integrated over the full response rather than taken from the degenerate-neutrino limit of Ref.~\cite{Reddy:1997yr}. In that limit, with the blocking of the final neutrino restored and $E=\mu_\nu$, our rate reproduces the degenerate result to 0.6\% at $T/E=1/300$. For a dilute gas of heavy nucleons it reproduces the static cross section $G_F^2E^2(h_V^2+3h_A^2)/\pi$ per nucleon to 0.3\%.

In the kinetic equation, Eq.~(3) of the main text, scattering is treated as isotropic and energy conserving, with the total rate $\sigma=c/\lambda_{\rm NC}$. This rate is computed without final-state neutrino blocking, because for energy-conserving scattering the blocking factors of the gain and loss terms cancel. The range column of Table~I of the main text includes a variation of $\sigma$ by a factor of two in either direction. Table~\ref{tab:S3} gives the scattering length at $E=T$, the energy used in the transparency criterion~\cite{Alford:2018lhf}. At $2n_0$ and 3~MeV it is 2.4~km on neutrons and protons, the circle in Fig.~1(a) of the main text, and 2.8~km on neutrons alone. At the capture energies the scattering length is much shorter, 93~m at 25~MeV for the same state, but the absorption length there is 5.0~m, so absorption remains the faster process.

\begin{table}[H]
\caption{Neutral-current scattering length, in km, of a neutrino with energy $E=T$ in DD2 matter, on neutrons and protons ($n+p$) and on neutrons alone ($n$).}
\label{tab:S3}
\begin{ruledtabular}
\begin{tabular}{lcccccc}
 & \multicolumn{2}{c}{$T=3$~MeV} & \multicolumn{2}{c}{$T=5$~MeV} & \multicolumn{2}{c}{$T=8$~MeV}\\
$n/n_0$ & $n+p$ & $n$ & $n+p$ & $n$ & $n+p$ & $n$\\
\hline
0.5 & 4.55 & 5.27 & 1.02 & 1.14 & 0.262 & 0.284\\
1 & 3.41 & 4.11 & 0.747 & 0.889 & 0.187 & 0.218\\
2 & 2.37 & 2.83 & 0.512 & 0.612 & 0.126 & 0.150\\
3 & 1.89 & 2.31 & 0.410 & 0.500 & 0.100 & 0.122\\
\end{tabular}
\end{ruledtabular}
\end{table}

\section{Kinetic response}
\label{sec:S4}

Each neutrino species $i=\nu,\bar\nu$ is described by its occupation $f_i(E,\mu,x,t)$, where $\mu$ is the cosine of the angle between the neutrino momentum and the wave vector. Absorption and emission give the collision term $j_i(1-f_i)-\kappa_if_i$, with the absorption rate $\kappa_i=c/\lambda_i$ and the emission rate $j_i$. Detailed balance, $j_i(1-f_{{\rm eq},i})=\kappa_if_{{\rm eq},i}$, turns it into $\gamma_i(f_{{\rm eq},i}-f_i)$ with the stimulated rate $\gamma_i=\kappa_i/(1-f_{{\rm eq},i})$. Linearised about $f_i^0$, this becomes $\gamma_i(\delta f_{{\rm eq},i}-\delta f_i)$, where $\delta f_{{\rm eq},i}=\mp f_i^0(1-f_i^0)\,\delta\mu/T$ follows from the neutrino chemical potentials $\mp\Delta\mu$. Neutral-current scattering adds $\sigma(\langle\delta f_i\rangle-\delta f_i)$ (Sec.~\ref{sec:S3}), and a plane wave $e^{i(kx-\omega t)}$ gives Eq.~(3) of the main text.

Dividing Eq.~(3) by $\gamma_i+\sigma-i\omega+ick\mu$ and averaging over directions gives $\langle\delta f_i\rangle=(\gamma_i\,\delta f_{{\rm eq},i}+\sigma\langle\delta f_i\rangle)A_i$, with
\begin{equation}
A_i=\frac12\int_{-1}^{1}\frac{d\mu}{\gamma_i+\sigma-i\omega+ick\mu}=\frac{1}{ck}\arctan\frac{ck}{\gamma_i+\sigma-i\omega},
\label{eq:Ai}
\end{equation}
and hence $\langle\delta f_i\rangle=\gamma_iA_i\,\delta f_{{\rm eq},i}/(1-\sigma A_i)$. The electrons gain one unit of lepton number for each $\nu_e$ absorbed or $\bar\nu_e$ emitted, and lose one for each $\nu_e$ emitted or $\bar\nu_e$ absorbed. Summing the collision terms over directions, energies and both species therefore gives $dq/dt=\Lambda\,\delta\mu$, with $\Lambda$ of Eq.~(4) of the main text.

With $\delta\mu=C\epsilon-Bq$ and $\epsilon\propto e^{-i\omega t}$, the composition follows from $-i\omega q=\Lambda\,\delta\mu$, so that $\delta\mu=-i\omega C\epsilon/(-i\omega+B\Lambda)$. The cycle-averaged power dissipated per volume, ${\rm Re}(\delta\mu^*\Lambda\,\delta\mu)/2={\rm Re}\,\Lambda\,|\delta\mu|^2/2$, equals the bulk-viscous power $\zeta\omega^2|\epsilon|^2/2$ for $\zeta=\zeta_{\rm eff}$ of Eq.~(5) of the main text. With the oscillation energy $K_T|\epsilon|^2/2$ per volume, as in Ref.~\cite{Alford:2019qtm} but with the isothermal $K_T$, the energy damping time is $\tau=K_T/(\zeta_{\rm eff}\omega^2)$.

At all 1800 combinations of the 45 DD2 states, nine wavelengths from 0.3 to 50~km together with $k\to0$, and four frequencies from 0.3 to 10~kHz, the admittance satisfies $0\le{\rm Re}\,\Lambda\le\Lambda_\infty$ and ${\rm Im}\,\Lambda\le0$. Because ${\rm Im}\,\Lambda\le0$, $|{-i\omega}+B\Lambda|^2\ge\omega^2+B^2({\rm Re}\,\Lambda)^2$, so that $\zeta_{\rm eff}\le C^2/(2B\omega)$. The kinetic damping time therefore cannot be shorter than $\tau_{\rm res}=2K_TB/(C^2\omega)$, the minimum of the transparent form at its resonance, and all computed damping times lie above this bound.

At long wavelengths the response separates into local relaxation and diffusion. Expanding Eq.~(\ref{eq:Ai}) to order $k^2$ gives
\begin{equation}
\Lambda(\omega,k)=\Lambda(\omega,0)+k^2\mathcal{D}(\omega)+O(k^4),\qquad
\mathcal{D}(\omega)=\sum_i\int dE\,\frac{c^2w_i\gamma_i^2}{3(\gamma_i+\sigma-i\omega)(\gamma_i-i\omega)^2},
\label{eq:smallk}
\end{equation}
where $\Lambda(\omega,0)$ is the closed-channel response of the main text~\cite{Sawyer:1980wp,Alford:2019kdw}. For $\omega\ll\gamma_i$, $\mathcal{D}$ reduces to $\sum_i\int dE\,c^2w_i/[3(\gamma_i+\sigma)]$, the diffusion of lepton number with the transport mean free path $c/(\gamma_i+\sigma)$. The resummed diffusion form
\begin{equation}
\Lambda_i\simeq\int dE\,w_i\gamma_i\,\frac{-i\omega+D_ik^2}{\gamma_i-i\omega+D_ik^2},\qquad D_i=\frac{c^2}{3(\gamma_i+\sigma-i\omega)},
\label{eq:P1}
\end{equation}
which has the same expansion, reproduces the damping time on all 45 DD2 states to 5.0\% at 10~km, 4.1\% at 20~km and 1.7\% at 50~km. At 10~km the closed-channel part supplies only 2--4\% of ${\rm Re}\,\Lambda$ at the five reference states, so the dissipation comes almost entirely from diffusion. Its contribution falls as $k^2$, which is why the damping time grows roughly as the square of the wavelength in Fig.~3(a) of the main text. Because $\mathcal{D}$ depends only weakly on $\omega$ for $\omega\ll\gamma_i$, the damping time of a 10-km mode changes by less than 14\% between 0.3 and 3~kHz (Table~\ref{tab:S9}).

The energy integrals use 128 Gauss--Legendre nodes on $0<E<50T$. With 256 nodes on $0<E<70T$, the damping times at the five reference states change by less than $10^{-5}$. The analytic angular average, Eq.~(\ref{eq:Ai}), agrees with a direct Gauss--Legendre quadrature over $\mu$ with 96 nodes to $10^{-12}$ in the damping time at 1~kHz and 10~km.

\section{Transparent baseline and the published map}
\label{sec:S5}

The transparent baseline describes matter from which all neutrinos escape. At fixed baryon density $n$, temperature $T$ and muon fraction $Y_\mu$, the electron fraction then changes as $dY_e/dt=F/n$, where $F(n,Y_e)=\Gamma_{\rm dec}-\Gamma_{\rm cap}$ is the difference between the decay and capture rates per unit volume, summed over direct and modified Urca. The rates are those of Secs.~\ref{sec:S2} and \ref{sec:S3}, evaluated at each composition with its own affinity, masses and occupations. The stationary electron fraction $Y_e^*$ solves $F(n,Y_e^*)=0$~\cite{Alford:2018lhf}. At the five reference states, $Y_e^*$ lies below the cold $\beta$-equilibrium value and the affinity at $Y_e^*$ is 11.3--17.5~MeV (Table~\ref{tab:S4}).

A density oscillation $\delta n=n\epsilon$ with $\epsilon\propto e^{-i\omega t}$ drives the composition. Linearising $dY_e/dt=F/n$ about $Y_e^*$ gives $\delta Y_e=\gamma_B\epsilon/(\gamma_I-i\omega)$, with $\gamma_I=-n^{-1}\partial F/\partial Y_e$ and $\gamma_B=\partial F/\partial n$. The out-of-phase part of the pressure then gives the bulk viscosity~\cite{Alford:2023gxq}
\begin{equation}
\zeta_{\rm tr}=-\left(\frac{\partial p}{\partial Y_e}\right)_{n,T,Y_\mu}\frac{\gamma_B}{\omega^2+\gamma_I^2}=\frac{nC\gamma_B}{\omega^2+\gamma_I^2},
\label{eq:zetatr}
\end{equation}
where the second form uses $C=-(\partial p/\partial n_e)_{n,T,Y_\mu}$ (Sec.~\ref{sec:S1}). The coefficients $C$ and $K_T$ are evaluated at $Y_e^*$, and $\tau_{\rm tr}=K_T/(\zeta_{\rm tr}\omega^2)$. If the rates depend on the composition only through the affinity, $F=\lambda\Delta\mu$, Eq.~(\ref{eq:zetatr}) reduces to the familiar form $\zeta=(C^2/B)\gamma/(\omega^2+\gamma^2)$ with $\gamma=B\lambda$~\cite{Alford:2018lhf,Alford:2019qtm}. The derivatives are taken by central differences with relative steps $2\times10^{-3}$ and $10^{-3}$, and the two steps give damping times that agree to $6\times10^{-4}$ on the whole grid. As stated in the main text, this baseline is not the short-wavelength limit of the kinetic model, which keeps the cold composition and a neutrino population in equilibrium with the matter.

\begin{table}[H]
\caption{Transparent baseline at the five reference states: cold $\beta$-equilibrium electron fraction $Y_e$, stationary electron fraction $Y_e^*$ with all neutrinos escaping, affinity $\Delta\mu^*$ at $Y_e^*$, the rates $\gamma_I$ and $\gamma_B$ of Eq.~(\ref{eq:zetatr}), and the damping time $\tau_{\rm tr}$ of a 1-kHz oscillation.}
\label{tab:S4}
\begin{ruledtabular}
\begin{tabular}{cccccccc}
$n/n_0$ & $T$ (MeV) & $Y_e$ & $Y_e^*$ & $\Delta\mu^*$ (MeV) & $\gamma_I$ (s$^{-1}$) & $\gamma_B$ (s$^{-1}$) & $\tau_{\rm tr}$ (ms)\\
\hline
0.5 & 3 & 0.0329 & 0.0251 & 11.3 & 1592 & 38.6 & 13.3\\
1 & 4 & 0.0507 & 0.0403 & 13.8 & 3324 & 98.4 & 19.4\\
1.5 & 4 & 0.0582 & 0.0488 & 12.7 & 1855 & 58.0 & 70.2\\
2 & 5 & 0.0633 & 0.0520 & 16.6 & 5784 & 177 & 69.7\\
3 & 5 & 0.0694 & 0.0587 & 17.5 & 5093 & 125 & 198\\
\end{tabular}
\end{ruledtabular}
\end{table}

The capped bars in Fig.~2 and the column Ref.~\cite{Alford:2019qtm} in Table~I of the main text are readings of the damping-time map for 1-kHz oscillations of DD2 matter in Fig.~6(a) of Ref.~\cite{Alford:2019qtm}. That map uses the saturation density $n_{\rm sat}=0.16$~fm$^{-3}$, whereas $n_0=0.149$~fm$^{-3}$ here, so we read it at equal physical density: a state at $xn_0$ lies at $0.932\,x$ on its density axis. We extracted the contour lines from the vector graphics of the published figure and give, for each state, the two contour values that bracket its temperature. The state at $0.5n_0$ lies below the lowest density of the map. Read at equal labels, $n/n_{\rm sat}=n/n_0$, the interval at $n_0$ and 4~MeV would be 50--100~ms, and the intervals at $1.5n_0$, $2n_0$ and $3n_0$ are unchanged. The map was computed with vacuum nucleon masses in the reaction rates, adiabatic susceptibilities and no muons~\cite{Alford:2019qtm}, so it serves as a reference interval rather than a direct comparison.

\section{Sensitivity to the rates}
\label{sec:S6}

Table~\ref{tab:S5} lists the damping times at the five reference states when one rate is rescaled at a time: the modified-Urca rate by a factor of three and the direct-Urca and neutral-current rates by a factor of two, in either direction. The range column of Table~I of the main text is the minimum and maximum over a mesh of $13\times5\times9=585$ combinations, with the modified-Urca factor between 1/3 and 3 and the direct-Urca and scattering factors between 1/2 and 2, each spaced uniformly in the logarithm. At every state the shortest damping time occurs for halved direct-Urca and scattering rates, and the longest for doubled scattering rates. From $n_0$ upward the longest also has the largest Urca rates, whereas at $0.5n_0$ it has the smallest.

Rescaling the Urca rates also changes the transparent baseline, through both its rates and its stationary composition $Y_e^*$. Recomputing the baseline with the same rescaled direct- and modified-Urca rates gives ratios $\tau_{\rm kin}/\tau_{\rm tr}$ of 25 to 540 for the four Urca variations at the five states, compared with 36 to 490 for the central rates. Neutral-current scattering does not enter the baseline.

\begin{table}[H]
\caption{Damping time (s) of a 1-kHz, 10-km oscillation at the five reference states for the central rates, with one rate rescaled at a time, and the minimum and maximum over the full mesh of rescaled rates.}
\label{tab:S5}
\begin{ruledtabular}
\begin{tabular}{cccccccccc}
 & & & \multicolumn{2}{c}{modified Urca} & \multicolumn{2}{c}{direct Urca} & \multicolumn{2}{c}{scattering} & \\
$n/n_0$ & $T$ (MeV) & central & $\times1/3$ & $\times3$ & $\times1/2$ & $\times2$ & $\times1/2$ & $\times2$ & mesh range\\
\hline
0.5 & 3 & 0.481 & 0.636 & 0.439 & 0.474 & 0.490 & 0.367 & 0.676 & 0.319--0.858\\
1 & 4 & 2.74 & 2.92 & 3.36 & 2.66 & 2.86 & 1.88 & 4.33 & 1.79--5.20\\
1.5 & 4 & 8.65 & 9.79 & 10.1 & 8.52 & 8.81 & 5.74 & 14.1 & 5.63--15.9\\
2 & 5 & 22.9 & 21.7 & 29.9 & 22.3 & 23.4 & 14.4 & 38.9 & 13.0--47.8\\
3 & 5 & 96.7 & 91.4 & 125 & 94.5 & 98.8 & 59.6 & 167 & 54.0--202\\
\end{tabular}
\end{ruledtabular}
\end{table}

\section{Other equations of state}
\label{sec:S7}

The IUFSU~\cite{Fattoyev:2010mx} and BigApple~\cite{Fattoyev:2020cws} EOSs are relativistic mean-field models with nonlinear meson couplings. We use their published parameters with the common nucleon mass of 939~MeV of the original fits, and their saturation density $n_0=0.155$~fm$^{-3}$ as the density unit. Each state is treated as for DD2: the cold $\beta$-equilibrium composition with the muon fraction held fixed, the finite-temperature affinity and coefficients at that composition, Landau masses from the cold Dirac masses, nucleon chemical potentials fixed by the densities at the temperature $T$ (Sec.~\ref{sec:S2}), and the transparent baseline of Sec.~\ref{sec:S5}. Table~\ref{tab:S6} lists the results. At $3n_0$ both EOSs lie within 16~MeV of the direct-Urca threshold, with $\Delta p=15.5$ and 8.9~MeV, where the rate treatment is not validated, and these states are not used in the main text. At $n_0$ the coefficient $C$ of BigApple is small, 6.9~MeV at the cold composition and 11.0~MeV at $Y_e^*$, which lengthens both of its damping times.

\begin{table}[H]
\caption{Damping times of a 1-kHz, 10-km oscillation for the IUFSU and BigApple EOSs at the states of Table~I of the main text: electron and muon fractions, momentum deficit $\Delta p$, affinity $\Delta\mu_0$, coefficient $C$, transparent baseline $\tau_{\rm tr}$, kinetic result $\tau_{\rm kin}$ with its range for rescaled rates (Sec.~\ref{sec:S6}), and the ratio $\tau_{\rm kin}/\tau_{\rm tr}$.}
\label{tab:S6}
\begin{ruledtabular}
\begin{tabular}{lccccccccccc}
EOS & $n/n_0$ & $T$ (MeV) & $Y_e$ & $Y_\mu$ & $\Delta p$ (MeV) & $\Delta\mu_0$ (MeV) & $C$ (MeV) & $\tau_{\rm tr}$ (ms) & $\tau_{\rm kin}$ (s) & range (s) & ratio\\
\hline
IUFSU & 0.5 & 3 & 0.0418 & 0.0000 & 75.9 & 1.83 & 15.9 & 6.54 & 0.558 & 0.397--0.925 & 85\\
 & 1 & 4 & 0.0501 & 0.0057 & 75.6 & 1.38 & 18.8 & 38.9 & 8.43 & 5.34--16.9 & 217\\
 & 1.5 & 4 & 0.0571 & 0.0181 & 62.7 & 0.75 & 46.0 & 36.9 & 4.90 & 3.16--9.41 & 133\\
 & 2 & 5 & 0.0653 & 0.0301 & 44.4 & 0.81 & 62.3 & 24.2 & 7.67 & 4.21--16.3 & 316\\
BigApple & 0.5 & 3 & 0.0477 & 0.0000 & 67.3 & 1.68 & 14.5 & 2.15 & 0.265 & 0.185--0.487 & 123\\
 & 1 & 4 & 0.0508 & 0.0061 & 74.0 & 1.36 & 6.9 & 129 & 64.4 & 40.6--129 & 501\\
 & 1.5 & 4 & 0.0567 & 0.0179 & 63.6 & 0.70 & 55.4 & 46.3 & 4.44 & 2.89--8.20 & 96\\
 & 2 & 5 & 0.0680 & 0.0322 & 38.3 & 0.70 & 81.6 & 28.0 & 7.37 & 4.29--15.1 & 264\\
\end{tabular}
\end{ruledtabular}
\end{table}

\section{Additional results}
\label{sec:S8}

Table~\ref{tab:S7} lists the kinetic and transparent damping times of a 1-kHz, 10-km oscillation on the full DD2 grid, including the densities $1.5n_0$ and $3n_0$, which are not shown in Fig.~2 of the main text. Tables~\ref{tab:S8} and \ref{tab:S9} list the dependence on wavelength and frequency at the two states of Fig.~3 of the main text.

The dashed curve in Fig.~1(a) of the main text is the elastic approximation used in opacity tables~\cite{Bruenn:1985en,Burrows:2004vq}. The nucleon momentum is unchanged, and the electron takes the energy $E_e=E+Q$ with $Q=m_n-m_p$, so that
\begin{equation}
\lambda_{\rm el}^{-1}=\sigma_0\,\frac{1+3g_A^2}{4}\left(\frac{E_e}{m_e}\right)^2\left(1-\frac{m_e^2}{E_e^2}\right)^{1/2}\left(1+1.1\,\frac{E}{m_n}\right)\eta_{np}\,\big[1-f_e(E_e)\big],
\label{eq:elastic}
\end{equation}
with $\sigma_0=1.705\times10^{-44}$~cm$^2$, the effective neutron density $\eta_{np}=(n_n-n_p)/\{1-\exp[(Q-\mu_n+\mu_p)/T]\}$, and the electron occupation $f_e$~\cite{Bruenn:1985en}. In this comparison the chemical potentials and the mean-field shift are taken from the cold background. The electron Fermi energy lies far above $E_e$ at the capture energies, so the factor $1-f_e$ suppresses this absorption. At $2n_0$ and 3~MeV it gives $\lambda_{\rm el}^{-1}=3\times10^{-20}$~m$^{-1}$ for a 25-MeV neutrino, or $3\times10^{-15}$~m$^{-1}$ when $Q$ includes the mean-field shift $U_n-U_p=30$~MeV~\cite{Oertel:2020pcg}, compared with 0.20~m$^{-1}$ from the full kinematics.

\begin{table}[H]
\caption{Kinetic damping time $\tau_{\rm kin}$ (s) and transparent baseline $\tau_{\rm tr}$ (ms) of a 1-kHz, 10-km oscillation in DD2 matter on the full grid of densities and temperatures.}
\label{tab:S7}
\begin{ruledtabular}
\begin{tabular}{ccccccccccc}
 & \multicolumn{2}{c}{$0.5n_0$} & \multicolumn{2}{c}{$n_0$} & \multicolumn{2}{c}{$1.5n_0$} & \multicolumn{2}{c}{$2n_0$} & \multicolumn{2}{c}{$3n_0$}\\
$T$ (MeV) & $\tau_{\rm kin}$ & $\tau_{\rm tr}$ & $\tau_{\rm kin}$ & $\tau_{\rm tr}$ & $\tau_{\rm kin}$ & $\tau_{\rm tr}$ & $\tau_{\rm kin}$ & $\tau_{\rm tr}$ & $\tau_{\rm kin}$ & $\tau_{\rm tr}$\\
\hline
2 & 1.40 & 210 & 9.69 & 2460 & 36.0 & 11900 & 86.9 & 27300 & 332 & 62700\\
2.5 & 0.674 & 42.2 & 4.48 & 408 & 16.0 & 1950 & 37.6 & 4680 & 145 & 12100\\
3 & 0.481 & 13.3 & 3.05 & 106 & 10.4 & 489 & 24.0 & 1210 & 94.3 & 3370\\
3.5 & 0.433 & 7.33 & 2.67 & 37.9 & 8.79 & 164 & 19.9 & 407 & 79.9 & 1200\\
4 & 0.450 & 8.79 & 2.74 & 19.4 & 8.65 & 70.2 & 19.3 & 171 & 79.6 & 516\\
4.5 & 0.503 & 16.0 & 3.08 & 16.8 & 9.31 & 41.7 & 20.6 & 91.3 & 86.2 & 275\\
5 & 0.575 & 30.9 & 3.60 & 23.3 & 10.5 & 38.5 & 22.9 & 69.7 & 96.7 & 198\\
6 & 0.749 & 99.3 & 4.95 & 68.8 & 14.0 & 78.0 & 29.5 & 107 & 125 & 255\\
8 & 1.22 & 652 & 8.36 & 485 & 23.6 & 447 & 47.7 & 515 & 195 & 1050\\
\end{tabular}
\end{ruledtabular}
\end{table}

\begin{table}[H]
\caption{Kinetic damping time (s) of a 1-kHz oscillation versus its wavelength at $n_0$, 4~MeV and $2n_0$, 5~MeV, and its limit without spatial transport, $k\to0$.}
\label{tab:S8}
\begin{ruledtabular}
\begin{tabular}{ccc}
Wavelength (km) & $n_0$, 4~MeV & $2n_0$, 5~MeV\\
\hline
0.3 & 0.0737 & 0.301\\
0.5 & 0.101 & 0.430\\
1 & 0.166 & 0.767\\
2 & 0.307 & 1.66\\
3 & 0.480 & 2.90\\
5 & 0.933 & 6.58\\
10 & 2.74 & 22.9\\
20 & 9.18 & 83.2\\
50 & 41.2 & 377\\
$k\to0$ & 141 & 1190\\
\end{tabular}
\end{ruledtabular}
\end{table}

\begin{table}[H]
\caption{Damping times (s) of a 10-km oscillation versus frequency at $n_0$, 4~MeV and $2n_0$, 5~MeV: kinetic result, short-wavelength limit of the kinetic model, and transparent baseline.}
\label{tab:S9}
\begin{ruledtabular}
\begin{tabular}{ccccccc}
 & \multicolumn{3}{c}{$n_0$, 4~MeV} & \multicolumn{3}{c}{$2n_0$, 5~MeV}\\
Frequency (kHz) & kinetic & short-wavelength & baseline & kinetic & short-wavelength & baseline\\
\hline
0.3 & 2.78 & 0.352 & 0.0623 & 23.2 & 3.27 & 0.393\\
1 & 2.74 & 0.0349 & 0.0194 & 22.9 & 0.301 & 0.0697\\
3 & 2.48 & 0.00702 & 0.0156 & 20.1 & 0.0393 & 0.0413\\
10 & 1.24 & 0.00384 & 0.0152 & 8.79 & 0.00956 & 0.0381\\
\end{tabular}
\end{ruledtabular}
\end{table}

\end{document}